\documentclass[prl,amsmath,amssymb,superscriptaddress,twocolumn]{revtex4}
\usepackage{graphicx}

\renewcommand{\Psi}{\varPsi}

\begin{document}

\title{Does string theory predict an open universe?}

\author{Roman~V.~Buniy} \email{roman@uoregon.edu}
\affiliation{Institute of Theoretical Science, University of Oregon,
Eugene, OR 97403}

\author{Stephen~D.~H.~Hsu} \email{hsu@uoregon.edu}
\affiliation{Institute of Theoretical Science, University of Oregon,
Eugene, OR 97403}

\author{A.~Zee}
\email{zee@itp.ucsb.edu}
\affiliation{Kavli Institute for Theoretical Physics,
University of California, Santa Barbara, CA 93106}

\begin{abstract}
It has been claimed that the string landscape predicts an open universe, with negative curvature. The prediction is a consequence of a large number of metastable string vacua, and the properties of the Coleman--De Luccia instanton which describes vacuum tunneling. We examine the robustness of this claim, which is of particular importance since it seems to be one of string theory's few claims to falsifiability. We find that, due to subleading tunneling processes, the prediction is sensitive to unknown properties of the landscape. Under plausible assumptions, universes like ours are as likely to be closed as open.
\end{abstract}



\maketitle

\section{Introduction} 

If, as suggested by recent results \cite{landscape}, string theory exhibits a landscape of more than $10^{500}$ distinct, metastable vacua, its status as a conventional scientific theory is in jeopardy. Most physicists feel that scientific theories must make predictions which are falsifiable by experiment. Such a large diversity of vacua -- the number may even be infinite -- might mean that essentially any low-energy physics is realizable from string theory. Optimistically, future work may reveal some testable properties of the landscape, such as coupling constant relations or constraints on particle content. However, at the moment we are unaware of any such predictions, with the possible exception of Weinberg's anthropic determination \cite{weinberg} of the value of the cosmological constant, and even that is sensitive to assumptions about the spectrum of primordial density perturbations \cite{ghjw}. Even ultra high-energy physics experiments may not yield additional information, since scattering at trans-Planckian energies leads to black holes \cite{BHP} of ever increasing size, whose subsequent behavior (evaporation) is controlled by the  low-energy physics of the ambient vacuum state. If recent results are any guide, string theory will be extremely difficult to falsify.

It is therefore important to carefully consider any robust implications of the string landscape, particularly those that might be testable in the forseable future. One of these, recently elaborated in \cite{Susskind}, is the prediction that our universe must be open, with negative curvature. A recent analysis combining WMAP and Sloan Digital Sky Survey data gives $\Omega_{\rm total} =1.003 \pm 0.010$ \cite{Omega}, but improved future observations could yield a statistically significant central value larger than unity, implying positive curvature. Would this rule out string theory?
 
The argument for an open universe is as follows. Consider an energy surface with many local minima, most of which have much more energy density than the observed value of the dark energy density $\Lambda \sim 10^{-10} {\rm eV}^4$. Given generic initial conditions, it is likely that our universe arrived at its current state via tunneling from a much more energetic metastable vacuum. Such tunneling processes are described by the Coleman-De Luccia (CDL) instanton \cite{CDL}, which exhibits an $O(4)$ symmetry in Euclidean space. 
For scalar fields, the instanton configuration with $O(4)$ symmetry has the lowest action \cite{CGM}. We assume that this is also the case when gravitational degrees of freedom are present. In that case, the Euclidean metric must have the form
\begin{equation}
\label{EM}
ds^2 = d\xi^2 + f(\xi)^2 d\Omega^2_{S^3}~,
\end{equation}
where $d\Omega^2_{S^3}$ denotes the distance element of a unit 3-sphere. The radial coordinate $\xi$ is orthogonal to families of 3-spheres satisfying $\sum_{i=1}^4 x_i^2 = \xi^2$. When this solution is analytically continued to Minkowski space, the corresponding metric  in the interior of the bubble is given by (\ref{EM}) for imaginary values of $\xi$, and with the 3-geometries $H^3$ satisfying the constraint
$$\sum_{i=1}^3 x_i^2- t^2 = r^2 - t^2 = \xi^2$$ ($t > r$ in the interior):
\begin{equation}
ds^2 = d\tau^2 - f(i \tau)^2 d\Omega^2_{H^3}~,
\end{equation}
where $\xi = + i \tau$. Here $H^3$ is defined to have signature $(--+)$. Following CDL, we have multiplied the metric by an overall sign in order to obtain their Minkowski signature. It is clear that the $H^3$ geometries are hyperbolic, and hence the resulting bubble universe is open.

We see that a resulting open universe depends crucially on the $O(4)$ invariance of the Euclidean solution, and the subsequent analytic continuation. In the next section we will investigate the properties of bubbles which might result from tunneling which is {\it not} dominated by a single Euclidean solution.

\begin{figure}[h]
\includegraphics[width=8cm]{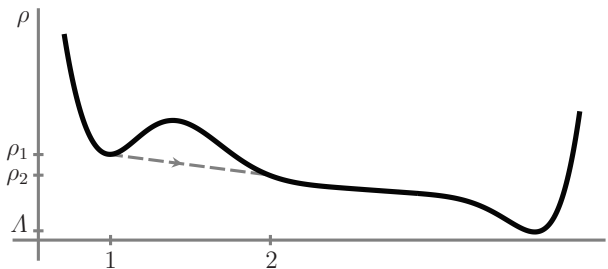}
\caption{A possible potential in the string theory landscape.}
\label{figure}
\end{figure}

In the remainder of this letter we will assume a potential of the form given in Figure 1. Tunneling takes place between configurations 1 and 2, with energy densities $\rho_{1,2}$ respectively. An extended flat region is assumed beyond the tunneling point 2, in order that the bubble interior experience an inflationary epoch after nucleation. In \cite{Susskind} it was deduced that roughly 60 e-foldings of inflation are necessary in order that the negative curvature of the CDL bubble not suppress galaxy formation.

For simplicity, we assume $\rho_1 \simeq \rho_2 = \Delta^4$, and a thin-walled bubble with surface tension $\sigma \sim \Delta^3$.
We define $$\epsilon \equiv {\rho_1 - \rho_2 \over \Delta^4} \ll 1~,$$ so a critical bubble has radius $r_* \sim (\Delta \epsilon)^{-1}$. This radius can be of order the de Sitter horizon size $r_{dS} \sim M / \Delta^2$ (where $M$ is the Planck scale) if $\epsilon$ is small and $\Delta$ not too small relative to $M$. This last condition is important for the non-Euclidean inflationary evolution of the interior of the bubble \cite{BGG}, as we will discuss further below.

\section{Beyond semiclassical dominance} 

It is easy to imagine situations in which the tunneling amplitude is not dominated by a single Euclidean solution. For example, if some of the fields involved in the tunneling are strongly coupled, the usual semiclassical expansion breaks down, and many paths in the functional integral play a role. Since there are so many field theory degrees of freedom on the landscape, it seems quite plausible that some of them will be strongly coupled. In addition, even if the leading amplitude is given by the CDL instanton, there will still exist subleading processes, albeit with exponentially smaller amplitudes. As explained below and in the following section, even exponentially suppressed bubbles are of interest, as they may play an important role in the anthropic calculation of probabilities.

We now relax the condition that the tunneling process be described by a Euclidean solution. We consider any amplitudes that are not explicitly forbidden by conservation laws, such as energy conservation. That is, any non-zero transition $\langle f \vert U \vert i \rangle$ where the initial state $\vert i \rangle$ is the vacuum 1 and the final state is that of a critical or supercritical (expanding) bubble of vacuum 2: $\vert f \rangle \equiv \vert B \rangle$.
We assume the bubble interior is homogeneous, so that deep inside its spacetime is described by the Friedman-Robertson-Walker (FRW) metric
\begin{equation}
ds^2 = dt^2 - R(t)^2 d\Omega_k^2~,
\end{equation} 
where the subscript $k$ on $d\Omega^2$, denotes the usual FRW open and closed geometries for negative and positive values, respectively. The bubble state $\vert B \rangle$, including metric and field theory degrees of freedom, is characterized by parameters such as $R, \dot{R}, \rho_2, \sigma, r $, with $r$ the bubble radius.

Given these assumptions, it is the initial conditions of the bubble that determine the curvature of the interior universe. Unlike in the CDL case, the evolution of the interior is not the analytic continuation of a particular Euclidean solution. Instead, we examine the Einstein equation for the interior
\begin{equation}
\label{E1}
H^2 = \kappa \rho - k / R^2~,
\end{equation}
where $H = \dot{R} / R$, and $\kappa = 8 \pi G / 3$. The sign of $k$ is determined by whether the initial interior density $\rho_2$ is greater or less than the critical density $\rho_c = H^2 / \kappa$:
\begin{equation}
k = \dot{R}^2 \left( \rho_2 / \rho_c - 1 \right)~.
\end{equation}
It is easy to arrange for positive $k$ without violating any conservation laws. For example, the bubble might resemble the CDL bubble, except with a smaller $\rho_c$ resulting from smaller initial $\dot{R}$. The total energy of the bubble can still be made equal to its volume times $\rho_1$, thereby conserving energy.

Eq.~(\ref{E1}) can be rewritten in the form of an energy conservation equation
\begin{equation}
\dot{R}^2 - \kappa \rho R^2 = - k Ê Ê.
\end{equation}
A region with $k>0$ can simply be regarded as one in
which the sum of negative gravitational binding energy and positive energy density in $\dot{R}$ is less than zero. Such a region is gravitationally bound: it decouples from the any
expansion (inflation) of the surrounding universe, and eventually
undergoes collapse. To an outside observer it ultimately appears as a black hole. Note that any region that is isotropic and homogeneous can be locally described by the FRW metric, and the value of k is simply determined by the rate of expansion relative to the matter energy density.

We mention an important caveat. We assume that the fate of the bubble interior can be deduced from FRW initial data describing it deep inside. In doing so, we neglect the boundary interactions of the expanding bubble wall with the false vacuum 2. In \cite{BGG} it was shown that inflationary (non-Euclidean) internal evolution of a {\it shrinking} false vacuum bubble is determined by a naive analysis of the interior properties if the bubble is sufficiently large, roughly the size of the de Sitter horizon determined by its interior energy density. In the previous section we noted that bubbles of this size are possible depending on the parameters $\epsilon$ and $\Delta$. In our case the bubble wall is expanding away from the interior at relativistic speed (i.e., the false vacuum is on the outside), whereas in the case studied in \cite{BGG} it is collapsing. In \cite{BGG}, when interior inflation is curtailed it is due to the impinging collapse, which is not an issue for our case. Our neglect of the boundary interaction is likely to be justified, particularly in the case of large bubbles.

In the remainder of the paper, we focus on the subset of bubbles $\vert B^* \rangle$ whose interiors are homogeneous and have $\rho$ very close to $\rho_c$ (i.e., $k$ very close to zero, with either sign). 
Such bubbles are particularly favorable for producing universes like ours, as first noted by Linde \cite{Linde}. Homogeneity ameliorates the horizon problem, while small initial $k$ lessens the flatness problem. The number of e-foldings of subsequent inflation required for such bubble universes to resemble ours is much less than the 60 or so required for the CDL bubble \cite{Susskind}. This opens the interesting possibility that, even if nucleation of $B^*$ bubbles is highly suppressed relative to CDL bubbles, the overall conditional probability that a universe {\it resembling ours} came from $B^*$ tunneling may be greater than or of order the probability that it came from CDL tunneling. The relative tunneling suppression might be compensated by the scarcity of flat potentials that can produce many e-foldings of slow roll inflation. 

Note that the background spacetime in which the $B^*$ bubble is nucleated has presumably been inflating for an extended period, and hence has almost exactly flat geometry. The curvature parameter $k$ of the $B^*$ bubble is assumed to be very close to zero, so we are essentially gluing a very slightly curved FRW region onto an exactly flat background. It seems unlikely that any boundary effects could cause a change in the sign of $k$.

\section{Anthropism and model dependence}  

Let us adopt the anthropic assumption that the landscape leads to the realization of many causally disconnected universes with varying properties. Further, let us assume that it is via bubble nucleation depicted in Figure 1 that universes with small cosmological constant are produced. Now, consider the conditional probability that a universe resembling ours (i.e., large, nearly flat, with structure formation) has $k$ either negative or positive. Whether positive or negative $k$ is more likely depends on whether the suppression of tunneling rates to $\vert B^* \rangle$ bubbles is compensated by less stringent conditions on the inflaton potential. For example, it seems plausible that the probability distribution for inflaton potentials (extended flat regions that allow slow roll de Sitter epochs) could be exponential in the number of e-foldings \cite{foot}. Then, even if $B^*$ tunneling is very rare relative to CDL events, there may be exponentially more vacua for which a $B^*$ bubble could lead to a universe like ours than for a CDL bubble.

In rough approximation, the ratio of the conditional probabilities is
\begin{equation}
\frac{ P( {\rm CDL} \vert {\rm us} ) }{ P( B^* \vert {\rm us} ) } ~\sim~
\frac{\Gamma({\rm CDL})}{\Gamma( B^* ) } \cdot  \frac {N ( 60^+)} {N ( B^* )}~,
\end{equation}
where $\Gamma ({\rm CDL} )$ and $\Gamma( B^* )$ denote tunneling rates, $N( 60^{+} )$ is the number of regions in the landscape of the type in Figure 1 which can produce at least 60 e-foldings, and $N(B^*)$ is the number that can produce a much smaller number of e-foldings sufficient to produce a universe like ours from $B^*$ initial conditions. (For example, a homogeneous $B^*$ bubble with $k$ close to zero might only require 40 e-foldings of inflation, and $N(B^*)$, the number of potentials that produce 40 e-folds of inflation might be exponentially larger than $N(60^+)$.)
We have assumed a large but finite number of vacua, so the $N$ are finite numbers. Even if 
$\Gamma ({\rm CDL} ) >> \Gamma( B^* )$, it could be that $N (B^*)$ is sufficiently larger than $N( 60^+ )$ that $P(B^* \vert {\rm us})$ dominates $P ( {\rm CDL} \vert {\rm us} )$. In that case, there could be roughly equal probabilities of $k$ slightly positive and negative, since $B^*$ bubbles could arise with either sign of the curvature. Even that conclusion is dependent on further assumptions, as the precise minimum amount of inflation necessary for $k$ negative and positive could be slightly different.

We might expect that if inflationary e-foldings are improbable, then the amount of curvature should be near an anthropic boundary.
What is the anthropic limit on flatness? Let us consider $k > 0$ 
since that is the case of most interest here.  We leave aside the dark
energy, which further complicates the issue. It is not clear
whether the dark energy will persist forever, e.g., as in the case
of a cosmological constant, or is only a temporary phenomenon 
(e.g., quintessence).

Positive curvature causes {\it less rapid} expansion than in a $k=0$
universe with similar matter content, so structure formation is not
the issue. Rather, the issue is whether the universe recollapses 
before life can evolve. Universes which survive for a long time before recollapse
might lead to {\it more} sentient beings (per unit volume) and be anthropically
favored. The number of beings might even grow exponentially with 
time, depending on population dynamics.  Therefore, even if
inflationary e-foldings are improbable, one cannot exclude $k > 0$
universes which are very flat. (Note, it would be quite surprising for the competing exponential pressures we describe to result in a spatial curvature which is observable but also consistent with current bounds!)

One might object that, if there is going to be exponential population 
growth in our future, then we are quite atypical of beings in our
universe. However, this objection requires a very strong use of the
typicality assumption which we do not find plausible. Used in this way, it would imply a dim future for the human race: if we are typical (i.e., not among the earliest humans to have lived), then the total future population of humans is quite limited! 

To conclude, it appears that the probability distribution for $k$ is dependent on detailed properties of the landscape and subtle anthropic assumptions \cite{foot}; the sign of cosmological curvature cannot be deduced by any simple arguments. Regrettably, it does not provide a clean test of string theory.

\smallskip

\begin{acknowledgments}
\section{acknowledgements}

S.~H. thanks F. Denef, M. Douglas and L. Susskind for discussions concerning the string landscape during the 2006 Ettore Majorana school of subnuclear physics.  A.~Z. thanks F. Wilczek for helpful discussions regarding the Coleman-De Luccia work. He is also grateful to the Radcliffe Institute for Advanced Study, Harvard University, where this work was done, for its warm hospitality. The authors thank anonymous referees for emphasizing the problems of eternal inflation, flat potentials and anthropic limits on curvature. S.~H. and R.~B. are supported by the
Department of Energy under DE-FG02-96ER40969, while A.~Z. is supported
in part by the National Science Foundation under grant number PHY
04-56556.

\end{acknowledgments}




\end{document}